\documentclass[preprint,showpacs,preprintnumbers,amsmath,amssymb,showkeys,floatfix,aps,prd,a4paper,byrevtex]{revtex4}

\usepackage{color}
\usepackage{epic}
\usepackage{eepic}
\usepackage{epsfig}
\usepackage{latexsym}
\usepackage{float}
\usepackage[all]{xy}
\usepackage{tikz} 
\usepackage[colorlinks, urlcolor=black, citecolor=red, linkcolor=blue]{hyperref}
\usepackage{graphicx}
\usepackage{dcolumn}
\usepackage{bm}
\usepackage{epsfig}
\usepackage{amsmath}
\usepackage{amsfonts}
\usepackage{amssymb,amscd}
\usepackage[latin1]{inputenc}
\usepackage{units}
\usepackage{hyperref}
\def\pom{{I\!\!P}}

\begin{document}


\title{Diffractive $Z$ boson pair production at LHC in the large extra dimensions scenario}

\author{V.~P. Gon\c{c}alves}
\email{barros@ufpel.edu.br}
\author{W.~K. Sauter}
\email{werner.sauter@ufpel.edu.br}
\author{M. Thiel}
\email{mauricio.thiel@gmail.com}
\affiliation{High and Medium Energy Group, \\
Instituto de F\'{\i}sica e Matem\'atica, Universidade Federal de Pelotas\\
Caixa Postal 354, CEP 96010-900, Pelotas, RS, Brazil}
\date{\today}

\begin{abstract}
In this paper we study the  diffractive  $Z$ boson pair production  mediated by the Kaluza-Klein graviton in the large extra dimensions scenario at the CERN Large Hadron Collider energies. Considering the Durham model we estimate the total cross section for the central inclusive and exclusive diffractive production of a $Z$ boson pair in $pp/pPb/PbPb$ collisions. Our results indicate that the  experimental analysis of the central inclusive production in $pp$ collisions is feasible at CERN LHC.
\end{abstract}

\maketitle

\section{Introduction \label{sec:intro}}

The search  for particles beyond the Standard Model (SM) is one of the key issues of the ATLAS and CMS  experiments at LHC. In particular, these experiments can test the theories with extra dimensions, which aim to solve the hierarchy problem by bringing the gravity scale closer to the electroweak scale (For a review see, e.g., \cite{hewett}). Over a decade ago,  Arkani-Hamed, Dimopoulos, and Dvali~\cite{add}
proposed a scenario whereby the  SM is constrained to the common 3 + 1 space-time dimensions (brane), while the gravity is free to propagate throughout a larger multidimensional space (bulk). 
In this Large Extra Dimensions (LED) scenario, $n$ extra spatial dimensions are compactified on a torus with common circumference $R$. Then the four dimensional Planck scale $M_P$ is no longer the relevant  scale but is related to the fundamental scale $M_S$ as follows $M_P^2 \approx M_S^{n+2}R^n$ where $M_S$ is ${\cal{O}}$(TeV).
Moreover, the $4+n$ dimensional graviton corresponds to a tower of massive Kaluza-Klein (KK) modes in four dimensions. The interactions of these spin-2 KK gravitons with the SM matter can be described by an effective theory with the  Lagrangian given by~\cite{giudice, han}
\begin{eqnarray}
{\cal{L}} = - \frac{\kappa}{2} \sum_{\vec{n}=0}^{\infty} T^{\mu \nu}(x) h_{\mu \nu}^{\vec{n}}(x) \,\,,
\label{lagrangiano}
\end{eqnarray} 
where $\kappa = \sqrt{16 \pi}/M_P$, the massive KK gravitons are labelled by a $d$-dimensional vector of positive integers and $T^{\mu \nu}$ denotes the energy-momentum tensor of the SM. 
In what follows we assume that 
\begin{eqnarray}
\kappa^2R^n = 8 \pi (4\pi)^{n/2} \Gamma(n/2)M_S^{-(n+2)}.
\end{eqnarray}
The Feynman rules that follow from Eq.~(\ref{lagrangiano}) can be found in Refs. \cite{giudice, han}. The individual KK resonances have masses equal to $m_{(\vec{n})} = |\vec{n}|/R$ and thus the mass gap between the neighbouring modes $\Delta m \propto 1/R$ is small for $n$ not too large, which allow us to approximate the discrete mass spectrum by a continuum.   
The individual KK mode couples with a strength $\propto 1/M_P$ to the SM fields. However, since there are many KK modes, the total coupling strength is of the order of $1/M_S$ after summing up all of them. Moreover, the excited gravitons preferentially decay into two gauge bosons rather than into two leptons, because the graviton has spin 2, and so fermions cannot be produced in an $s$ wave.

\begin{figure}[t]
\begin{tabular}{cc}
\includegraphics[scale=0.35] {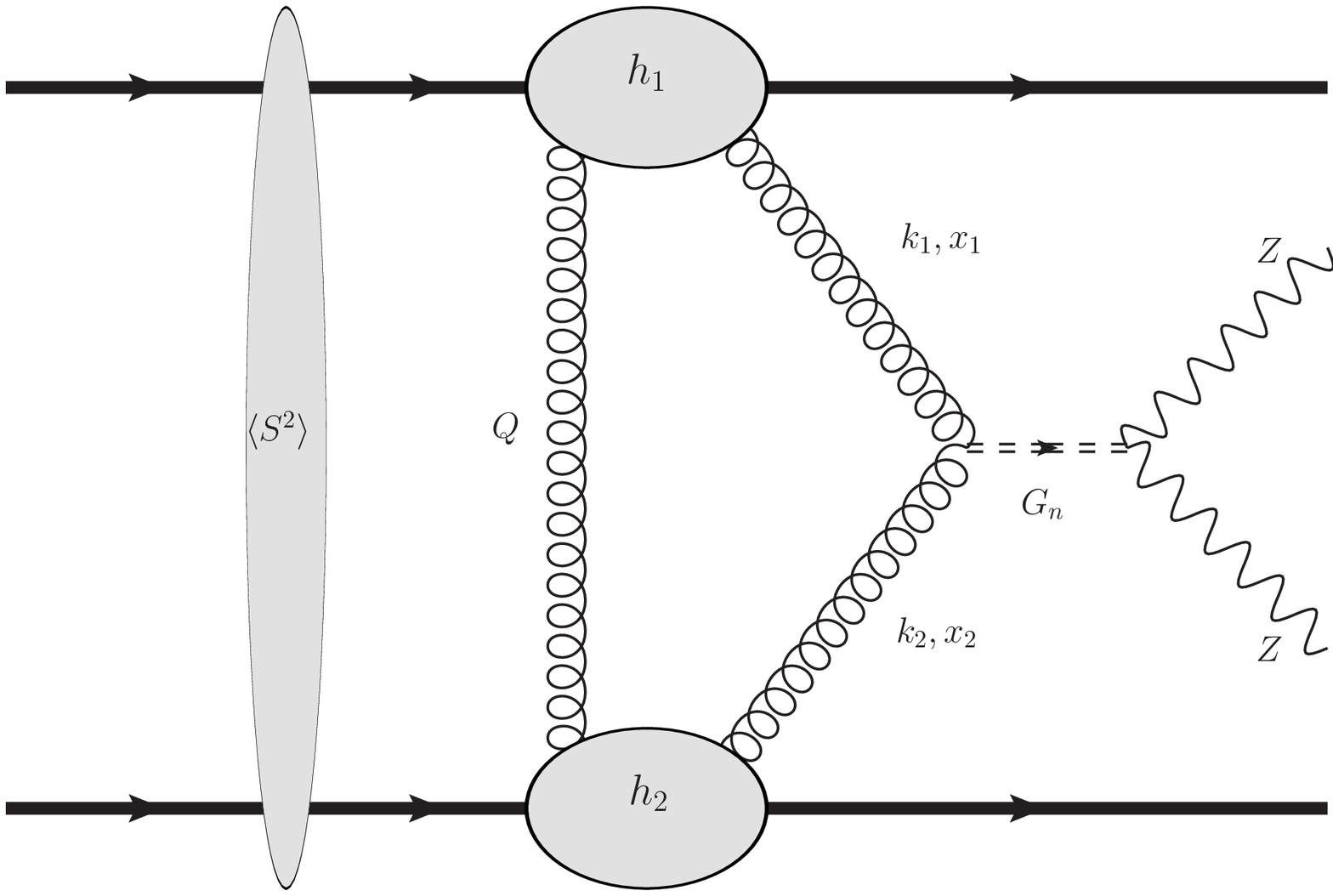} & \includegraphics[scale=0.35]{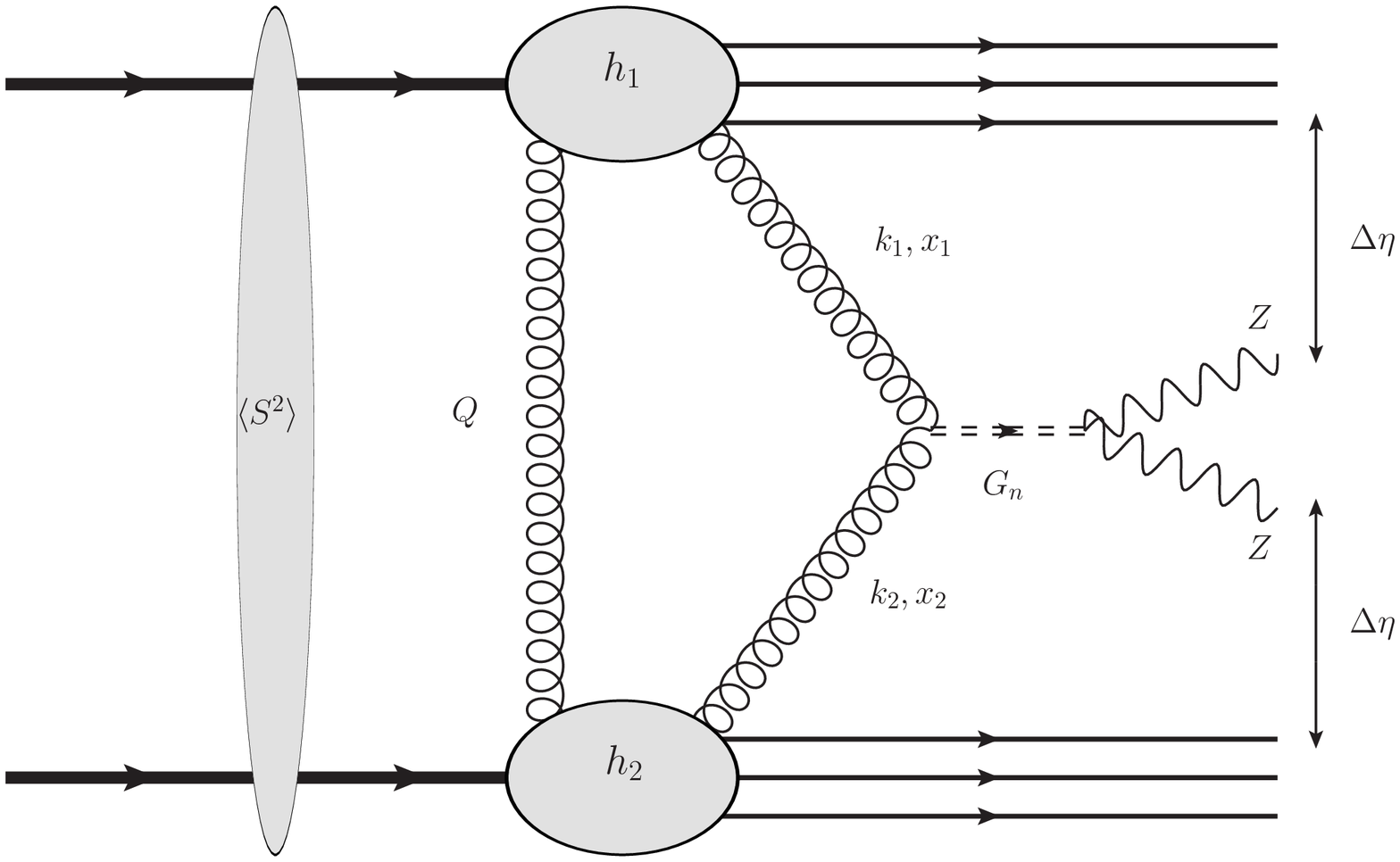} \\
(a) & (b)
\end{tabular}
\caption{(color online) $Z$ boson pair production in (a) central exclusive and (b) central inclusive diffractive processes  for  $h_1 h_2$ collisions, with $h_i = p$ or $Pb$. $\langle {\mathcal S}^2 \rangle$ is the survival probability gap, which gives the probability that secondaries, which are produced by soft rescatterings, do not populate the rapidity gaps.}
\label{fig1}
\end{figure}

Searches for the large extra dimensions scenario via virtual-graviton effects  were performed at HERA, LEP, the Tevatron, and the LHC (For a recent review see, e.g., \cite{review_exp}). Recent experimental results from the ATLAS and CMS Collaborations sets lower limits on $M_S$, being about 3.5 TeV (2.7 TeV) for $n = 2$ ($n = 6$) \cite{refsexp}. From the theoretical side, many studies on the virtual KK graviton effects up to the NLO exist ~\cite{calculos_nlo}. These include the processes of fermion-pair, multijet, and diboson production. 
In general the contribution associated to the graviton exchange is mostly undetectable because of the much larger Standard Model background. The lower SM background occurs in the  production of $Z$ pairs which have the smallest cross section of all the diboson processes. In contrast, this process have tree-level contributions in the LED scenario, which  motivated the study of  the  $Z$ boson pair production  mediated by the Kaluza-Klein graviton in large extra dimensions \cite{koch,gao,agarwal,ravindran_recent}. These studies consider inclusive  processes, where the hadrons colliding dissociate after the interaction. In this paper we extend these previous studies for {\it diffractive} processes, in which the hadrons colliding  lose only a small fraction of their initial energy and escape the central detectors~\cite{forshaw_review}.   In the diffractive interaction the hadron can dissociate (inclusive diffractive process) or remain intact (exclusive diffractive process).  In what follows we consider the 
$Z$ boson pair production in central exclusive processes (CEP), $h_1 + h_2 \rightarrow h_1 \otimes ZZ \otimes h_2$, which are characterized by  empty regions in pseudo-rapidity, called rapidity gaps (denoted by $\otimes$), that separate the intact very forward hadron from the  $ZZ$ final state produced in the interaction [See Fig.~\ref{fig1}(a)]. Exclusivity means that nothing else is produced except the leading hadrons and the central object. Moreover, we consider central inclusive processes (CIP), $h_1 + h_2 \rightarrow X \otimes ZZ \otimes Y$, which also exhibit rapidity gaps but the incident hadrons dissociates in the products $X$ and $Y$ [See Fig.~\ref{fig1}(b)]. In general the rapidity gaps present in inclusive processes are smaller than in the exclusive case. A QCD mechanism for the central inclusive and exclusive diffractive production of a heavy central system has been proposed by Khoze, Martin and Ryskin~\cite{kmr_first,kmr_prosp,kmr_review}, denoted Durham model hereafter,  with its predictions in 
broad agreement with the observed rates measured by CDF Collaboration~\cite{cdf} (For a recent review see Ref.~\cite{forshaw_review}). In this paper  the Durham model is applied, for the first time, for the $Z$ boson pair production considering the KK graviton exchange. Our main motivation is associated to the fact that in the Standard Model  the $Z$ boson pair is not produced at leading order by the gluon fusion. Consequently, the $Z$ boson pair cannot be produced at leading order in diffractive processes. At next-to-leading-order the $gg \rightarrow ZZ$ subprocess via the quark loop contributes \cite{zzsm}. However, as will be demonstrated below for the case of central inclusive processes, the resulting cross section is  one order of magnitude smaller than the LED prediction. It implies  a  very low QCD background, making its observation a signature of  large extra dimensions.

This paper is organized as follows. In the next section we present a
brief review of the formalism necessary for the calculation of the $Z$ boson pair   production in diffractive interactions in hadron - hadron collisions. Moreover, we discuss the extension of the survival probability gap for nuclear collisions. In Section~\ref{sec:res}  we present our results for the $Z$ boson pair production in $pp$, $pPb$ and $PbPb$ collisions at LHC energies. Finally, in Section~\ref{sec:sum} we present a summary of our main conclusions.

\section{The $Z$ boson pair production in diffractive interactions \label{sec:modfor}}

In the Durham model \cite{kmr_prosp} the total cross section for the diffractive production of a $Z$ boson pair reads
\begin{equation}
 \sigma(\sqrt{s}) =   \int dy \int \frac{dM^2_{ZZ}}{M^2_{ZZ} }{\mathcal L}_{gg} \hat{\sigma}_{gg \rightarrow ZZ}(M^2_{ZZ})
 \label{sigma}
\end{equation}
where  ${\mathcal L}_{gg}$ is the effective gluon - gluon luminosity for the production of a system with squared invariant mass $M^2_{ZZ}$ and rapidity $y$,  and $\hat{\sigma}$ is the cross section for the subprocess $gg \rightarrow ZZ$. 
For the $Z$ boson pair production in central exclusive processes (CEP), $h_1 + h_2 \rightarrow h_1 \otimes ZZ \otimes h_2$, the effective gluon - gluon luminosity is given by [See Fig.~\ref{fig1}(a)]
\begin{equation}
 {\mathcal L}_{gg}^\mathrm{excl} =  \langle {\mathcal S}^2_{excl} \rangle \left[ {\cal{C}} \int \frac{dQ_t^2}{Q_t^4} f_g(x_1,x_1^{\prime},Q_t^2, \mu^2) f_g(x_2,x_2^{\prime},Q_t^2, \mu^2)\right]^2\,\,,
\end{equation}
where $\langle {\mathcal S}^2_{excl} \rangle$ is the survival probability gap for exclusive processes, which gives the probability that secondaries, which are produced by soft rescatterings, do not populate the rapidity gaps and  ${\cal{C}} = \pi/[(N_C^2 - 1)b]$, with $b$ the $t$-slope ($b = \unit[4.0]{GeV^{-2}}$ in what follows). Moreover,  $Q_t^2$ and $x_i^{\prime}$ are respectively the virtuality and longitudinal momentum of the soft gluon needed for color  screening, $x_1$ and $x_2$ are the longitudinal momentum of the gluons which participate of the hard subprocess   and the quantities $f_g$ are the  skewed unintegrated gluon densities.  Since $
(x^{\prime} \approx {Q_t}/{\sqrt{s}}) \ll (x \approx {M_{ZZ}}/{\sqrt{s}}) \ll 1$,
 it is possible to express $f_g(x_1,x_1^{\prime},Q_t^2, \mu^2)$, to single log accuracy, in terms of the conventional integral gluon density $g(x)$, together with a known Sudakov suppression $T$ which ensures that the active gluons do not radiate in the evolution from $Q_t$ up to the hard scale $\mu = 2 M_{Z}$, which we assume to be the renormalization and factorization scales of the process.  The choice of this scale introduces roughly a factor of two uncertainty when varying the hard scale $\mu$ between $2 M_{ZZ}$ and $M_{ZZ}/2$.  Following~\cite{kmr_prosp} we will assume that
\begin{equation}
 f_g(x, Q_t^2, \mu^2) = S_g \frac{\partial}{\partial \ln Q_t^2} \left[ \sqrt{T(Q_t,\mu)}\ xg(x,Q_t^2) \right]\,\,,
\end{equation}
where $S_g$ accounts for the single $\log Q^2$ skewed effect and is given by
\[ S_g = \frac{2^{2\lambda+3}}{\sqrt{\pi}} \frac{\Gamma(\lambda + 5/2)}{\Gamma(\lambda + 4)} \]
if one assumes a single powe-law behaviour for the gluon distribution, $xg(x,Q_t^2) \approx x^{-\lambda}$, with $S_g \sim 1.2 $ for LHC. The Sudakov factor is given by
\begin{eqnarray}
T(Q_t,\mu)  =  \exp \left\{ -\int_{Q_t^2}^{\mu^2} \frac{dk_t^2}{k_t^2} \frac{\alpha_s(k_t^2)}{2\pi} \int_{0}^{1-\Delta} dz \, \left[ zP_{gg}(z) + \sum_{q} P_{qg}(z) \right] \right\},
\end{eqnarray} 
with $k_t$ being an intermediate scale between $Q_t$ and $\mu$, $\Delta = k_t/(\mu + k_t)$, and $P_{gg}(z)$ and $P_{qg}(z)$ are the leading order Dokshitzer - Gribov - Lipatov - Altarelli - Parisi (DGLAP) splitting functions~\cite{dglap}. 
In this paper we will calculate  $f_g$ in the proton case considering that  the integrated gluon distribution $xg(x,Q_T^2)$ is described by the MSTW2008lo parametrization~\cite{mstw}. In the nuclear case we will include the shadowing effects in $f_g^A$ considering that the nuclear gluon distribution is given by the EKS98 parametrization~\cite{eks}, where  $ x g_A(x,Q_t^2) = A R_g^A(x, Q_t^2) x g_p(x, Q_t^2)$, 
with $R_g$ describing the nuclear  effects in $xg_A$. 
In order to obtain realistic predictions for the exclusive $Z$ boson pair production  using the Durham model, it is crucial to use an adequate value for the survival probability gap, $\langle {\mathcal S}^2_{excl} \rangle$. This factor is the probability that secondaries, which are produced by soft rescatterings do not populate the rapidity gaps (For a detailed discussion see~\cite{kmr_review}). In the case of proton-proton collisions, we will assume that $\langle {\mathcal S}^2_{excl} \rangle = 3 \, \%$
for LHC  energies~\cite{kmr_prosp}. However, the value of the survival probability for nuclear collisions still is an open question. An estimate of  $\langle {\mathcal S}^2_{excl} \rangle$ for nuclear collisions was calculated in \cite{miller} using the Glauber model. Another conservative estimate can be obtained assuming that \cite{vicwerner}:  $\langle {\mathcal S}^2_{excl} \rangle_{A_1A_2} = \langle {\mathcal S}^2_{excl} \rangle_{pp}/(A_1 \cdot A_2)$ (For a discussion about nuclear diffraction see~\cite{dif_nuc}). In what follows we will consider the latter model for $\langle {\mathcal S}^2_{excl} \rangle_{A_1A_2}$ when considering the central exclusive $Z$ boson pair production.

On the other  hand, the effective gluon - gluon luminosity for the $Z$ boson pair production in central inclusive processes (CIP), $h_1 + h_2 \rightarrow X \otimes ZZ \otimes Y$, is given by~\cite{kmr_prosp} 
\begin{eqnarray}
 \mathcal{L}_{gg}^\mathrm{incl} = & \langle {\mathcal S}^2_\mathrm{incl} \rangle \frac{\alpha_s^4}{\pi}\left(\frac{N_C^2}{N_C^2 -1}\right)^2 \frac{1}{(Y_1 + Y_ 2)^2}\int \frac{dt_1}{t_1}\frac{dt_2}{t_2}\exp\left(-\frac{3\alpha_s}{\pi}\Delta\eta\left|\ln\frac{t_1}{t_2}\right|\right) \times \nonumber \\ 
 & T(\sqrt{|t_1|}, {\mu}) T(\sqrt{|t_2|}, {\mu}) \int^1_{x_1^\mathrm{min}}\frac{dx_1}{x_1}\mathcal{G}(x_1, k^2_{1\perp}) \int^1_{x_2^\mathrm{min}} \frac{dx_2}{x_2}\mathcal{G}(x_2, k^2_{2\perp}),
\end{eqnarray}
where $\langle {\mathcal S}^2_\mathrm{incl} \rangle$ is the gap survival probability for inclusive processes, $t_i = - k^2_{i\perp}$, $\Delta\eta$ is the rapidity gap size on either side of the $Z$ boson pair and $Y_1 = Y_2 = \frac{3\alpha_s}{2\pi}\Delta\eta$, with $\alpha_s = 0.2$  [See Fig.~\ref{fig1}(b)]. Moreover,  $\mathcal{G}(x_i, k^2_{i\perp}) $ are the effective parton distributions given by
\begin{equation}
 \mathcal{G}(x_i, k^2_{i\perp}) = x_i g(x_i, k^2_{i\perp}) + \frac{N_c^2}{(N_c^2 -1)^2}\sum_{q} x_i\left[ q(x_i, k^2_{i\perp}) +\bar{q}(x_i, k^2_{i\perp}) \right].
\end{equation} 
and  
\[ x_i^\mathrm{min} = \frac{e^y}{\sqrt{s}} \left[ M_{ZZ} + k_{i\perp}e^{\Delta\eta} \right]. \]
The basic idea is that the initial hadrons dissociate and the system $ZZ$ with mass $M_{ZZ}$ is produced with rapidity gaps with size $\Delta \eta$ on either side. This process is characterized by two rapidity gaps, the central system $ZZ$ and the presence of secondary particles in the hadron fragmentation regions. An open question is the magnitude of the absorption corrections for the central inclusive processes which determines $\langle \mathcal{S}^2_\mathrm{incl} \rangle$, which is expected to depend on the kinematics of the process~\cite{kmr_prosp}. As it is expected that $\langle {\mathcal S}^2_\mathrm{excl} \rangle \le \langle \mathcal{S}^2_\mathrm{incl} \rangle$, we will assume, for simplicity, in our analysis that $\langle \mathcal{S}^2_\mathrm{incl}\rangle_{pp}  = 0.1$. Moreover, we will also assume that  $\langle {\mathcal S}^2_\mathrm{incl} \rangle_{A_1A_2} = \langle {\mathcal S}^2_\mathrm{incl} \rangle_{pp}/(A_1\cdot A_2)$. However, this subject deserve more detailed studies in the future.

In contrast to the Standard Model, in the LED scenario, the $gg \rightarrow ZZ$ subprocess  contributes at the tree level through the exchange of virtual KK gravitons in the $s$-channel, with the cross section being given by~\cite{gao},
\begin{equation}
 \hat{\sigma}_{gg \rightarrow ZZ} = \int_{-1}^{+1} d(\cos \theta)  \frac{d\hat{\sigma}_{gg}}{d\cos\theta} \,\,,
\end{equation}
with
\begin{eqnarray}
\frac{d\hat{\sigma}_{gg}}{d\cos \theta}&=&\frac{
\pi\hat{s}^3\sqrt{1-4w}
\mathcal{F}^2(n)\left(\pi^2+4\mathcal{T}^2(n)\right)}{1024}\nonumber\\
&&\times\left\{
3z^4\left(1-4w\right)^2+2z^2\left(1-4w\right)\left(5+12w\right)
+\left(1+12w\right)\left(3+4w\right)\right\},
\end{eqnarray} 
and
\begin{equation}
{\mathcal F}(n)=\hat {s}^{n/2-1}/M_S^{n+2}, \ {\mathcal
T}(n)=I(\Lambda/\sqrt{\hat s}),
\end{equation}
where $w=m_Z^2/\hat{s}$ and $z=\cos\theta$.
The function $I(\Lambda/\sqrt{\hat s})$ is directly related with the graviton propagator in the LED scenario~\cite{han}, being given by 
\begin{equation}
 I(\Lambda/\sqrt{s}) = PV\int_0^{\Lambda/\sqrt{s}} dy \frac{y^{n-1}}{1-y^2} \,\,,
\end{equation}
where $PV$ means the principal value of the integral and $\Lambda$ is identified with the fundamental scale $M_S$ in $4+n$ dimensions~\cite{giudice,han}.


\section{Results \label{sec:res}}

Before to present our predictions for the central  exclusive and inclusive $Z$ boson pair production in diffractive processes, a comment is in order. Our calculations are performed at leading order. Next-to-leading-order (NLO) corrections for the inclusive $Z$ boson pair production, $p + p \rightarrow ZZ X$, where both colliding protons dissociate and rapidity gaps are not present in the final state, were estimated in Ref.~\cite{agarwal}
assuming that the interaction is  mediated by the Kaluza-Klein graviton in the LED scenario 
  and  are large ($\approx \, 2$). Consequently, our predictions for the inclusive production should be considered a lower bound. In contrast, the magnitude of the NLO corrections for diffractive processes still is an open question (For a recent study, see Ref.~\cite{forshawcou}). Results presented in Refs. \cite{cudell,dechambre} indicate that the predictions for central production are strongly  influenced by the choice of the parametrization used for the parton distribution functions and by the treatment of the Sudakov form factor and rapidity gap survival probability, which introduces non-negligible uncertainties in the calculations. Such uncertainties also are present in our analysis. In principle, the uncertainties could be reduced with a possible early LHC measurement of exclusive jets cross section \cite{dechambre}.

\begin{figure}[t]
\begin{center}
\includegraphics*[width=8cm]{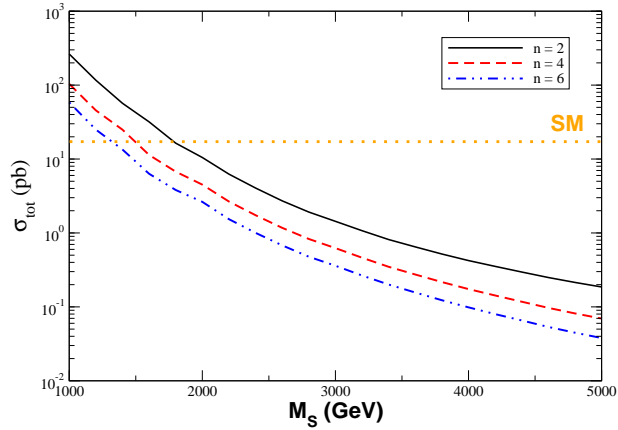} 
\caption{(color online) Dependence on $M_S$ of the cross section for the inclusive $Z$ boson pair production in $pp$ collisions at $\sqrt{s} = 14$ TeV. The SM prediction also is presented for comparison.}
\label{fig2}
\end{center}
\end{figure} 

In Fig.~\ref{fig2} we present our predictions for the inclusive production ($p + p \rightarrow ZZ X$) at $\sqrt{s} = \unit[14]{TeV}$.   For comparison we also show the SM prediction for this process, which is $\sigma_{tot} = \unit[17]{pb}$ and was obtained at NLO in Ref.~\cite{zzsm} (For recent experimental results see, e.g. Ref.~\cite{Chatrchyan:2013oev}).   As already verified in Ref.~\cite{gao}, the $Z$ boson pair production by KK graviton exchange is only competitive for small values of $M_S$ and number of extra dimensions $n$. In particular, if we taken into account the recent lower bounds on $M_S$ set by the ATLAS and CMS Collaborations \cite{refsexp}, we have that the LED predictions are at least a factor 20 smaller than the SM one.

\begin{figure}[t]
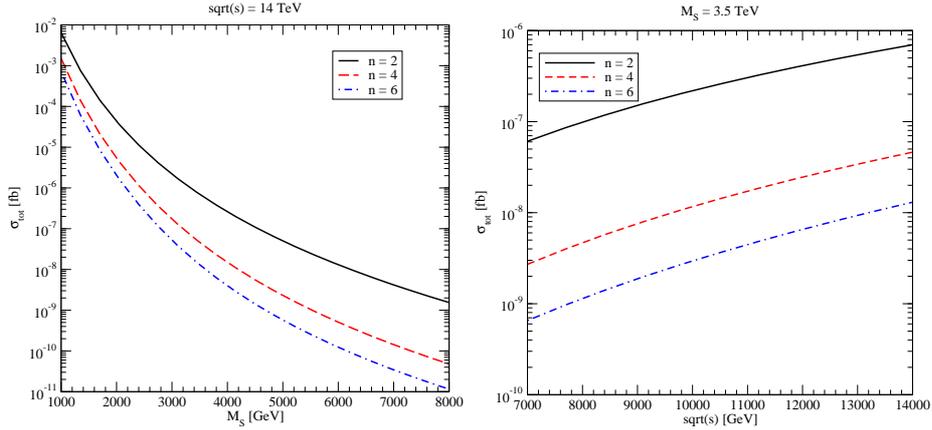

\begin{center}
\begin{tabular}{cc}
 \includegraphics*[width=6cm]{zz_excl_ms-n.eps} & \includegraphics*[width=6cm]{zz_excl_sqrts-n_REV.eps} \\
\end{tabular}
\caption{(color online) Dependence on  $M_S$  (left panel) and on the center-of-mass energy  (right panel) of the total cross section for the central {\it exclusive} double $Z$  production in  proton-proton collisions for different numbers of extra dimensions.}
\label{fig:exclpp}
\end{center}
\end{figure} 

In Fig.~\ref{fig:exclpp} we present our predictions for the central exclusive production of a $Z$ boson pair, $p + p \rightarrow p \otimes ZZ \otimes p$,  at LHC energies. In the left panel we present the dependence on $M_S$ of the total cross section for different values of $n$ at fixed center-of-mass energy ($\sqrt{s} = \unit[14]{TeV}$). As expected, it  strongly decreases at larger values of $M_S$ and $n$. In the right panel, we present our predictions for the energy dependence of the cross section considering $M_S = \unit[3.5]{TeV}$ and different values of $n$. We predict very small values for the exclusive cross section in the LHC kinematical range, which makes the experimental analysis of this process a very hard task.

\begin{figure}[ht]
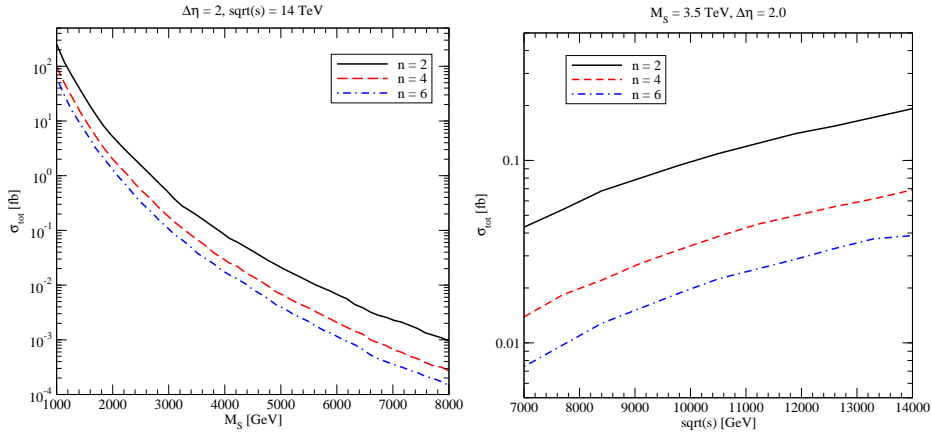

\begin{center}
\begin{tabular}{cc}
 \includegraphics*[width=6cm]{zz_incl_ms-n.eps} & \includegraphics*[width=6cm]{zz_incl_sqrts-n_REV.eps} \\
\end{tabular}
\caption{(color online) Dependence on $M_S$ (left panel) and on the center-of-mass energy (right panel) of the total cross section for the central {\it inclusive} double $Z$  production in  proton-proton collisions for different numbers of extra dimensions.}
\label{fig:incpp}
\end{center}

\end{figure}

\begin{figure}[t]
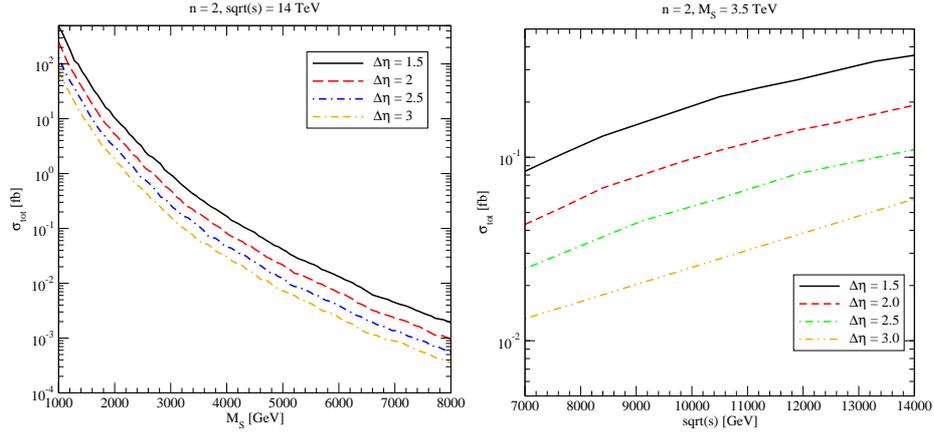

\begin{center}
\begin{tabular}{cc}
 \includegraphics*[width=6cm]{zz_incl_ms-eta.eps} & \includegraphics*[width=6cm]{zz_incl_sqrts-eta_REV.eps} \\
\end{tabular}
\caption{(color online) Dependence on $M_S$ (left panel) and on the center-of-mass energy  (right panel) of the total cross section for the central {\it inclusive} double $Z$  production in  proton-proton collisions for different values of the rapidity gap size $\Delta \eta$.}
\label{fig:incpp2}
\end{center}

\end{figure}

In Fig.~\ref{fig:incpp} we present our predictions for the central inclusive production of a $Z$ boson pair, $p + p \rightarrow X \otimes ZZ \otimes Y$, at LHC energies obtained from Eq.~(\ref{sigma}) assuming  that $|y| \leq 2$ and $M_{ZZ} \le M_S - \unit[10]{GeV}$. Moreover, initially we assume a fixed value for the rapidity gap size on either side of the $Z$ boson pair: $\Delta \eta = 2$. As in the exclusive case, the cross section decreases at larger values of $M_S$ and $n$. However, the magnitude is a factor $10^3$ larger, which becomes feasible its experimental study. In particular, we predict values of the order of fb at $\sqrt{s} = \unit[14]{TeV}$,  $M_S = 3.5$ TeV and $n = 2$. In Fig.~\ref{fig:incpp2} we present the dependence of our predictions on the rapidity size $\Delta \eta$. The cross section decreases by a factor $2.5$ when $\Delta \eta$ increases from 2.0  to 3.0.  Consequently, our predictions are not strongly modified by the choice of $\Delta \eta$. 

\begin{figure}[t]
\begin{center}
\includegraphics*[width=8cm]{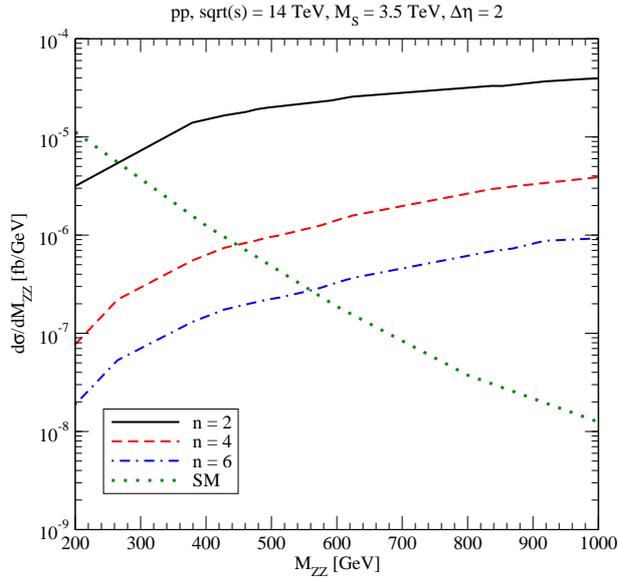} 
\caption{(color online) $Z$ pair invariant mass $M_{ZZ}$ distribution of the  central inclusive  cross section  in $pp$ collisions at $\sqrt{s} = 14$ TeV. The SM prediction also is presented for comparison.}
\label{fig:mass}
\end{center}
\end{figure} 

In Fig. \ref{fig:mass} we present our predictions for the $Z$ pair invariant mass $M_{ZZ}$ distribution of the  central inclusive  cross section  in $pp$ collisions at $\sqrt{s} = 14$ TeV. 
For comparison we also present the diffractive SM prediction which consider the next-to-leading-order  $gg \rightarrow ZZ$ subprocess via a quark loop diagram \cite{zzsm}. The LED predictions dominate the distribution at larger values of the invariant mass, as expected from previous studies for the inclusive production of different final states \cite{calculos_nlo}. In particular, we obtain that the LED prediction for the  total cross section in central inclusive processes is one order of magnitude larger than the SM one for $M_S = 3.5$ TeV and $n = 2$.

\begin{figure}[t]
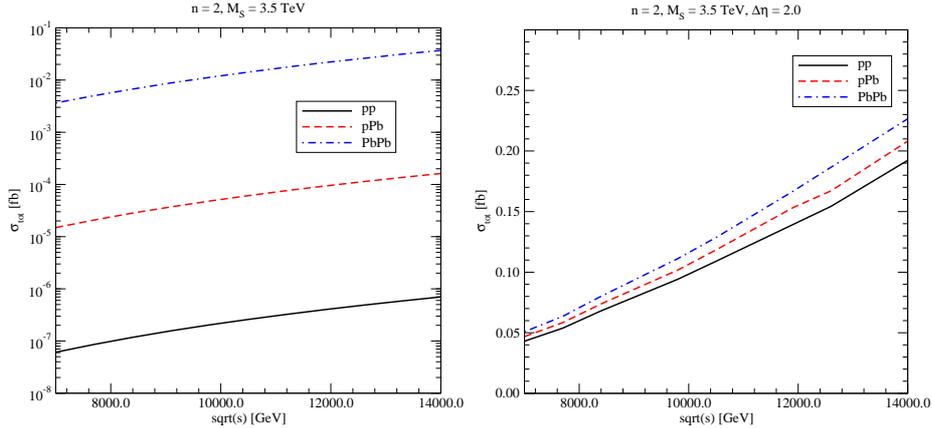

\begin{center}
\begin{tabular}{cc}
 \includegraphics*[width=6cm]{zzexclnuclear_REV.eps} & \includegraphics*[width=6cm]{zzinclnuclear_REV.eps} \\
\end{tabular}
\caption{Double $Z$ production for central  exclusive (left panel) and inclusive (right panel) processes in $pp/pPb/PbPb$ collisions.}
\label{fig:nuclear}
\end{center}

\end{figure}

Let us now present our predictions for the double $Z$ production in $pPb$ and $PbPb$ collisions at LHC energies. In Fig.~\ref{fig:nuclear} we present our predictions for the energy dependence of the central exclusive (left panel) and inclusive (right panel) cross section assuming $n = 2$, $M_S = \unit[3.5]{TeV}$ and $\Delta \eta = 2$. For comparison the $pp$ predictions also are shown. We have verified that  the $M_S$, $n$, $\Delta \eta$ and $M_{ZZ}$ dependences for the nuclear cross sections are very similar to the $pp$ one. In contrast,  the magnitude of the cross sections are very different for $pp$, $pPb$ and $PbPb$ collisions. In particular, for central exclusive processes the $PbPb$ cross section is amplified by a factor $\approx 10^4$ in comparison to the $pp$ one. It is directly associated to the strong dependence of the cross section on the nuclear gluon distribution, $\sigma^{A_1A_2}_{excl} \propto [xg_{A_1} \times xg_{A_2}]^2$, where    $x g_{A_i}= A_i R_g^{A_i}\cdot x g_p$. Consequently, if we consider 
our 
conservative  model for the nuclear survival probability  $\langle {\mathcal S}^2_\mathrm{excl} \rangle_{A_1A_2}$ we obtain that $\sigma^{A_1A_2}_{excl} \propto A_1 R_g^{A_1} \times A_2 R_g^{A_2} \times \sigma^{pp}_{excl}$, with $R_g > 1$ in the kinematical range probed by the double $Z$ production. On the other hand, in central inclusive processes the difference between the predictions is smaller, since in this case $\sigma^{A_1A_2}_{incl} \propto  R_g^{A_1} \cdot  R_g^{A_2} \times \sigma^{pp}_{incl}$. It implies that the predictions for the  central exclusive and inclusive productions are of the same order in $PbPb$ collisions.  For $pPb$ collisions, the central inclusive production dominates.

\begin{table}
\begin{center}
\begin{tabular} {||c|c|c||}
\hline
\hline
& {\bf CEP} &  {\bf CIP}  \\
\hline
\hline
$pp$ ($\sqrt{s} = \unit[14]{TeV}$) &  $8.0  \times 10^{-5}$  & $19$  \\
\hline
$pPb$ ($\sqrt{s} = \unit[8.8]{TeV}$) & $2.7 \times 10^{-8}$ & $3.1 \times 10^{-5}$  \\
 & $1.8 \times 10^{-5}$ & 0.02 \\
\hline
$PbPb$ ($\sqrt{s} = \unit[5.5]{TeV}$) & $2 \times 10^{-8}$ & $0.1 \times 10^{-6}$ \\
\hline
\hline
\end{tabular}
\end{center}
\caption{ The events rate/year for the  double $Z$ production  in $pp/pPb/PbPb$  collisions at LHC energies considering the CEP and CIP mechanisms. We assume $n=2$, $M_S = \unit[3.5]{TeV}$ and $\Delta \eta$ = 2. }
\label{tabZZ}
\end{table}

Finally, now we compute the production rates for LHC energies  considering the distinct mechanisms.  We assume $n=2$, $M_S = \unit[3.5]{TeV}$ and $\Delta \eta$ = 2.  The results are presented in Table~\ref{tabZZ}. At LHC we assume the  design luminosities ${\cal L} = 10^7 /\, 150 /\, \unit[0.5]{mb^{-1}s^{-1}}$ for $pp/pPb/PbPb$  collisions at $\sqrt{s} = 14/\,8.8/\,\unit[5.5]{TeV}$ and a run time of $\unit[10^7 \, (10^6)]{s}$ for collisions with protons (ions). 
Moreover, we also consider the upgraded $pPb$ scenario proposed in Ref.~\cite{david}, which analyse a potential path to improve the  $pPb$ luminosity and the running time. These authors proposed the following scenario for $pPb$ collisions:   ${\cal L} = \unit[10^4]{mb^{-1}s^{-1}}$  and a run time of $\unit[10^7]{s}$. The corresponding event rates are presented in the third line of the  Table~\ref{tabZZ}.
Our results indicate that for the default settings and running times, the statistics are marginal for $PbPb$ collisions.  Consequently, the possibility to carry out a measurement of the double $Z$ production in diffractive interactions is virtually null in these collisions. On the other hand, in $pp$ collisions the event rates for the central inclusive production predicted by the LED scenario is reasonable, with the SM background being one order of magnitude smaller. In comparison to the results for the inclusive double $Z$ production presented in Fig.~\ref{fig1} and  Ref.~\cite{gao}, our predictions  for diffractive production are a factor $\approx 10^{4}$  smaller.  Despite their much smaller cross sections, the clean topology of the diffractive production implies a larger signal to background ratio. Therefore, the experimental detection is in principle feasible. However, the signal is expected to be reduced due to the event pileup which eliminates one of the main advantages  of the diffractive processes.
In contrast, in $pA$ collisions it is expected to trigger on and carry out the measurement with almost no pileup~\cite{david}. Therefore, the upgraded $pA$ scenario also provides one possibility to detect the double $Z$ production in central inclusive processes.

\section{Summary \label{sec:sum}}
The search for new physics at the TeV-scale is one of the major tasks for the current and future high-energy physics experiments. In particular, models with extra spatial dimensions and TeV scale gravity are expected to provide a plethora of new and interesting signals. In this paper the double $Z$ production in diffractive processes  was considered for the first time. As in the Standard Model  the $Z$ boson pair is not produced at leading order by the gluon fusion, this final state cannot be produced at leading order in diffractive processes, which implies that a  very low QCD background is expected. We have estimated the cross sections and event rates for the double $Z$ production in central exclusive and central inclusive processes considering $pp/pPb/PbPb$ collisions. Both processes are characterized by the presence of two rapidity gaps in the final state, differing by the dissociation or not of the incident hadrons.	Our results indicate that the LED predictions for the central inclusive production are larger than the SM one and that  
the  experimental analysis of this process in $pp$ collisions is feasible at CERN LHC. This conclusion motivates a more detailed analysis of the gap survival probability for the double $Z$ production, which is one of the main sources of uncertainty in our calculations, as well as to implement our calculations in a Monte Carlo simulation for diffractive processes. Both subjects will be addressed in the future.
\section*{Acknowledgements}
This work was partially financed by the Brazilian funding agencies CNPq, CAPES and FAPERGS.


\end{document}